\documentclass[12pt]{article}
\newfont{\feff}{cmti10}
\usepackage{graphicx}
\usepackage{subfigure}
\usepackage{epstopdf}

\begin{document}

\title{Reynolds Number of Transition   as a Dynamic Constraint on Statistical Theory of Turbulence. }
\author{Victor Yakhot\\
Department of  Mechanical Engineering,\\ Boston
University, Boston MA 02215 }

\maketitle

\begin{abstract}
\noindent    Iterative coarse-graining procedure based on  Wyld's perturbation expansion is applied to the problem of  Navier-Stokes turbulence.  
It is shown that  the low-order calculation gives the  fixed-point Reynolds number $ Re_{fp}$  (coupling constant)   almost identical 
to  the Reynolds number of   the recently discovered transition to anomalous scaling of the moments of {\bf ``velocity derivatives''}.  
Using this result as a dynamic constraint, 
it is argued  that  in the vicinity of the fixed point  (integral scale)  the high-order non-linearities,  generated by the procedure,   are irrelevant .  
The infra-red divergencies do not disappear but are are contained in the derived 
equations for the symmetry-breaking   large-scale flows (turbulence models or ``condensates''),   which are    source of the small-scale turbulence.


\end{abstract}

{\it Introduction.}     For many years   patterns,  emerging in fluids undergoing  transition  to turbulence,   were   a source of  fascination not only to scientists but also to artists and philosophers.  Further increase of   the Reynolds number  leads to  a complex system of intermittently bursting and dying   small-scale structures  resembling ``worms'', rolls or ``pancakes''.   Still, even when the Reynolds number is very large, one can discern  somewhat blurred,  but   clearly seen  with naked eye    silhouettes  of   large-scale images  created at the transition point Re=$Re_{tr}$.  An example, which is part of our every-day experience,  is  ``vortex street'' behind a fast car moving on a dusty  road.  These patterns can be  made  clearer when small-scale fluctuations  are filtered out. 

\noindent Disregarding intermittency, a turbulent flow is characterized by two length scales:  integral  $L\approx 1/\Lambda_{f}$ where energy is pumped into the system and dissipation scale $\eta=1/\Lambda_{0}$ at which viscous effects balance the non-linearity.  According to Kolmogorov's theory, neither $L$ nor $\eta$ can appear in the expression for the energy spectrum in the ``inertial range''.  If this is so, the energy spectrum must be $E(k)\propto {\cal E}^{\frac{2}{3}}k^{-\frac{5}{3}}$. Due to 
infra-red divergencies of renormalized perturbation expansions  of  turbulence theory,  this qualitatively appealing result has never been derived directly from the Navier-Stokes equations .  It is well-known that each term in Wyld's   diagrammatic expansion is infra-red (i.r.) divergent, i.e. tends to infinity together with integral scale $L\approx 1/\Lambda_{f}$ [1].    Among the earliest  attempts to ``save''  the theory was Kraichnan's Lagrangian History Direct Interaction Approximation (LHDIA)   which was basically a one-loop closure  written in Lagrangian coordinates   eliminating  the effects of a transport of small eddies by the large ones [2].  Still, the theory was unable to deal with subleading  contributions appearing in the higher  orders. Moreover,  according to  moderns experimental and numerical data, it is quite possible that, due to  anomalous scaling,  the integral scale  does enter  the expressions for  moments of velocity increments in  the inertial - range scales   $1/\Lambda_{0}\ll r\ll L$.

\noindent   Application of the dynamic renormalization group to statistical theory of fluids was first proposed in Refs. [3]-[4].  These ideas were later generalized to the 
problem of large-scale features of hydrodynamic turbulence in Refs.[5]-[7].  
In the lowest order in powers of the turbulent- viscosity-based Reynolds number  (the $\epsilon$-expansion) the renormalization group (RNG)   led to an excellent 
agreement with experimental data on various dimensionless amplitudes characterizing large-scale features of turbulent flows [5]-[7]. 
 Moreover, the method  yielded the coarse-grained equations ( turbulence models)  widely used in modern engineering [8]-[9]. 
The main drawback of  the theory can be illustrated  as  follows.  According to Kolmogorov's phenomenology  the scale-dependent    {\bf coupling constant}  (Reynolds number) $Re(r)=u(r)_{rms}r/\nu(r)$, where $\nu(r)$ is the effective turbulent viscosity accounting  for the effects of the small-scale fluctuations from the interval $l<\leq r$. 
In the inertial range neither integral nor dissipation scale $\eta$ can enter the result.  Therefore $\nu(r)\approx u_{rms}(r)r$ and at the the fixed point of renormalization group $Re(r)\approx O(1)=const$ is not a small parameter.  Technically speaking, this invalidates the low-order truncation of expansion in powers of effective Reynolds number and the reasons  for numerical success of the theory  are  yet to be explored.  To correct for this drawback, one has to use non-perturbative methods  or resummations  of an infinite series in powers of the $O(1)$ parameters. \\

\noindent It will be shown below  that the Reynolds number $Re(r)$ derived at the fixed point  in Refs. [5]-[7] is almost identical  to that of {\it smooth''} transition to strong turbulence numerically discovered Ref.[10].   Below, using this result as a dynamic constraint on an expansion,   we argue that  at the fixed point  $r=1/\Lambda_{f}$ all high-order nonlinear terms (HOT),  generated by  coarse - graining,  sum up to zero,  thus justifying the $\epsilon$- expansion introduced in [5]-[7].  It will also become clear that outside the fixed point, in the inertial range,  the HOT exponentially grow invalidating some of the  theories based on the  second-order closures. \\

{\it Transition to turbulence.}   For almost a century  
transition to turbulence   has been  a  major theoretical challenge.  There exist  a huge literature on this topic  which,  together with theory of dynamical systems, evolved into a separate field of research.  Typically, one searches for instabilities  in  laminar flows  manifested by  an exponential growth of some modes ${\bf u(k,}t)$. 
We will loosely  identify   laminar flow as  a pattern ${\bf u}_{0}$  formed by a small set of excited modes  supported in the range of wave-numbers $k\approx  \Lambda_{f}$.
 All modes with $k>\Lambda_{f}$ are strongly overdamped, i.e. $u(k)=0$  for both $k\ll \Lambda_{f}$ and $k\ll \Lambda_{f}$. \\

\noindent  {\it Landau's theory.}  Here we mention just one work which is relevant for  considerations presented below. Assuming that in the vicinity of a transition point imaginary part of complex frequency is much smaller than the real one,  Landau considered the Navier-Stokes equations for incompressible fluid.
Denoting the velocity field at a transition point   ${\bf u}_{0}$ and introducing an infinitesimal perturbation ${\bf u}_{1}$ he wrote  ${\bf u}={\bf u}_{0}+{\bf u}_{1}$ with ${\bf u}_{1}=A(t)f({\bf r})$. Based the on qualitative considerations,  Landau  proposed [10]:

\begin{eqnarray}
\frac{d|A|^{2}}{dt}=2\gamma|A|^{2}-\alpha |A|^{4}\nonumber
\end{eqnarray}

\noindent where in the vicinity of transition point $\gamma=c(Re-Re_{tr})$ and $\alpha>0$.   In principle, $|A|^{2}$ must be considered as time- averaged.  Landau noted, however, that ${\bf u}_{1}({\bf k})$ is a slow mode and, since the averaging is taken over relatively short time -intervals, the averaging  sign in the above equations  is not necessary.  
At small times the solution exponentially grows and then reaches the maximum 
 $A_{max}\propto \sqrt{Re-Re_{tr}}$.  When  $\gamma=Re-Re_{tr}<0$, any initial perturbation decays.  In  this theory, 
 the magnitude of transitional Reynolds number is a free parameter and since the  large-scale field ${\bf u}_{0}$ strongly depends on geometry, external forces, stresses   
 the transition Reynolds number  $Re_{tr}$ 
 is not expected to be a universal constant.
 




\noindent  Landau assumed that  further increase of the Reynolds number leads to instability of first unstable mode generating next  excited modes  with the wave-vectors $k_{2}>k_{1}$ etc.  In  the modern lingo, this process can be perceived as an onset  of the energy cascade toward small- scale excitations with $k>\Lambda_{f}$.  
This leads  to formation of ``inertial range'' and  strongly intermittent  small-scale dissipation rate ${\cal E}$. \\

\noindent  {\it Transition to turbulence: a new angle.}  A new way  of looking at  phenomenon of transition to turbulence was  introduced  in numerical simulations  of a flow   at a relatively low Reynolds number $R_{\lambda}=\sqrt{\frac{5}{3{\cal E}\nu}}u_{rms}^{2}\geq 4.0$ [11]. 
In this work the   rms velocity  was defined as: $u_{rms}=\sqrt{u_{x}^{2}+u_{y}^{2}+u_{j}^{2}}$ 
which included not only turbulent fluctuations at $Re>Re_{tr}$ but also a low-Reynolds number  (``non-turbulent'') velocity field  ${\bf u}_{0}$ at $Re=Re_{tr}$. 
 In this approach transition to turbulence is identified with  the first appearance of non-gaussian anomalous  fluctuations of velocity derivatives including those of dissipation rate.
The flow in a periodic box was generated by a force in the right-side of the Navier-Stokes equation
with driven by the force
${\bf F}({\bf k},t)={\cal P}\frac{{\bf u}({\bf k},t)}{\sum' |\bf {u}({\bf k},t)|^{2}}\delta_{\bf k,k'}
$, where summation is carried over ${\bf k}_{f}=(1,1,2); \  (1,2,2)$. It is easy to see that the model with this forcing   generates flows with constant energy flux ${\cal P}={\cal E}=\overline{\nu(\frac{\partial u_{i}}{\partial x_{j}})^{2}}=const$ and the variation of the Reynolds number is achieved by  
variation of viscosity.  

\noindent    The results of Ref.[11] can be briefly summarized as  follows: 1. \ Extremely well-resolved simulations of the low-Reynolds number  flows   at $R_{\lambda}\geq 9-10$ revealed a clear scaling range  $\overline{(\frac{\partial u}{\partial x})^{n}}\propto Re^{\rho_{n}}$ with the anomalous scaling exponents $\rho_{n}$ consistent with the inertial range exponents typically observed only in  very high Reynolds number flows $Re\gg Re_{tr}$.  Identical scaling exponents $\rho_{n}$  were later obtained in some other flows [12]  indicating possibility of a broad universality. 
2. \ For  $R_{\lambda}<9-10$ the flow was  subgaussian indicating a dynamical system 
consisting of a small number of modes with  the small-scale fluctuations strongly  overdamped. 
 This flow  can be called ``quasilaminar'' or coherent.   3. \ At  a transition point $R_{\lambda,tr}\approx 9-10$ the fluctuating  velocity derivatives obey gaussian statistics and at $R_{\lambda}>9- 10$ a strongly anomalous scaling of the moments, typical of high-Reynolds number turbulence,  is clearly seen.   4. \ It has also been noticed that transition is smooth, i.e.  velocity field at ${\bf u}(R_{\lambda,tr}^{-})-{\bf u}(R_{\lambda,tr}^{+})\rightarrow 0$.  \\
 
 \noindent Below, we consider turbulence driven by a  force ${\bf F}(\Lambda_{f})$  supported at  the range of scales $\Lambda_{f}$.    
 Like in Landau theory, let us denote  the velocity field at the very onset of turbulence ${\bf u}_{0}$ and $Re_{tr}=\frac{u_{0,rms}}{\nu_{tr}\Lambda_{f}}$.  Keeping forcing fixed, by decreasing viscosity one achieves a large Reynolds number flow. Neglecting for a time being (see below) eddy noise (backscattering) we assume that the integral scale $\Lambda_{f}$ is Reynolds number independent and the principle feature of strong turbulence is formation of small-scale fluctuations in the range $k>\Lambda_{f}$ . In this case the velocity field $\overline{{\bf u}}_{0}\approx const$. 
 In this picture, the  large-scale patterns,  created  in the vicinity of a transition point $Re_{tr}$,  stay unchanged  but somewhat blurred by   random gaussian corrections which will be considered in detail below.
 The velocity field in the large-Reynolds number turbulent flow is then  ${\bf U}={\bf u}_{0}+{\bf u}$ so that $Re=Re_{tr}\frac{\nu_{tr}}{\nu} + \frac{u_{rms}}{\Lambda_{f}\nu}$ where $\nu\ll \nu_{tr}$.

\noindent An interesting feature of turbulent flows deserves discussion. According to experimental data,  when $r<<1/\Lambda_{f}$, the 
probability density of velocity increments $\Delta_{r}u_{i}=u_{i}(x+r)-u_{i}(x)$ leads  to the moments $S_{n}(r)=\overline{(\Delta_{r}u)^{n}}$  characterized 
by anomalous scaling  $S_{n}\propto r^{\xi_{n}}$.  However, in the limit $r\rightarrow \Lambda_{f}$ the moments obey gaussian statistics disregarding geometric 
features of the large-scale flow pattern.  This quite general observation has not yet been discussed in the literature.\\

\noindent{\it The model.}   Stirring  a very viscous fluid with the force ${\bf F}$ supported on a scale $r\approx 1/\Lambda_{f}$  leads to formation of a laminar flow field, with velocity ${\bf u(k)}$ often proportional to ${\bf F}(k)$. Gradually reducing viscosity to the magnitude $\nu=\nu_{tr}$ one reaches a transition point of marginal stability of a  laminar pattern. Upon further decrease of viscosity the flow becomes totally unstable which is accompanied by generation of small-scale modes ${\bf u(k)}$ with $k>\Lambda_{f}$ where $F(k)=0$.  In the limit $\nu\rightarrow 0$, the so called inertial range is formed with velocity fluctuations $u(k)$ filling  the interval $\Lambda_{f}\leq k\leq \Lambda_{0}\rightarrow \infty$.  According to this picture, turbulent flow is an interplay of the marginally stable , somewhat ``noisy'', velocity field ${\bf u}_{0}$ formed at transition  with small-scale fluctuations ${\bf u}(k)$ with $k>\Lambda_{f}$. The integral scale $\Lambda_{f}$ corresponding to the top of the inertial range is an essential  part of dynamics which is to be dealt with. \\

\noindent  Based on  these  qualitative features, we consider a flow generated by the Navier-Stokes equations with  a  force ${\bf F}(\Lambda_{f})$ in the right side:

\begin{equation}
\frac{\partial {\bf u}}{\partial t}+{\bf u\cdot \nabla u}=-{\nabla}p +{\bf \nu_{0}\nabla^{2}{\bf u}+ f+{\bf F}(\Lambda_{f})};  
\end{equation}

\noindent  with  $\nabla\cdot{\bf u}=\nabla\cdot{\bf f}=\nabla\cdot{\bf u}=0$. If $f=0$ and $\nu=\nu_{tr}$ , the model (1) represents a quasilaminar flow ${\bf u}_{0}$ at a transition point.  The random force, 
mimicking small-scale fluctuations  is defined by the correlation function:

\begin{equation}
\overline{f_{i}({\bf k},\omega)f_{j}({\bf k'},\omega')}=2D_{0}(2\pi)^{d+1}k^{-y}P_{ij}({\bf k})\delta({\bf k+k'})\delta(\omega+\omega');\hspace{1cm} f_{i}(k\leq \Lambda_{f},t)=0.
\end{equation}

\noindent  on the interval  $\Lambda_{f}< k\leq \Lambda_{0}$, so that $\overline{f_{i}F_{j}}=0$.   The i.r. cut off $\Lambda_{f}$, corresponding to the yet unknown top of the inertial range,  must be expressed in terms of  observables  like energy  ${\cal K}=u^{2}_{rms}/2$ and dissipation rate ${\cal E}$, respectively.  It is clear  that,  in a statistically steady state production must be equal to dissipation  so that ${\cal P}={\cal E}=\nu_{0}\overline{(\frac{\partial u_{i}}{\partial x_{j}})}$   and:

\begin{equation}
{\cal E}={\cal P}=\overline{{\bf F\cdot u}}+2D_{0}\frac{S_{d}}{(2\pi)^{d}}\frac{\Lambda_{f}^{-y+d}-\Lambda_{0}^{-y+d}}{y-d}
\end{equation}

\noindent The turbulent field  ${\bf u}$ is independent on the forcing so that  $\overline{{\bf F\cdot u}}= \overline{{\bf F\cdot u_{0}}}$. In the limit $y>d$, the information about the u.v. cut-off $\Lambda_{0}$ disappears and we can conclude that in the range $y>d$, the model (1)-(3) may lead  to universal statistics of velocity fluctuations in the inertial range of scales.  As $y\rightarrow d$, the energy flux and mean dissipation rate 
$ {\cal P}\rightarrow \overline{{\bf F\cdot u}}+2D_{0}\frac{S_{d}}{(2\pi)^{d}}\ln\frac{\Lambda_{0}}{\Lambda_{f}}=const$
 shows logarithmic dependence on the u.v. cut-off.    In the limit $\Lambda_{0}\rightarrow \Lambda_{f}$, the turbulent component disappears  and the balance between production and dissipation is given by ${\cal E}=\overline{{\bf F\cdot u}}$. \\
 
\noindent When $Re>Re_{tr}$ one can introduce ``turbulent''  Reynolds number: $Re_{T}=u_{rms}/(\nu\Lambda_{f})$,  and, if the energy spectrum $E(k)\propto k^{-\alpha}$, turbulent kinetic energy is defined  as:

\begin{equation}
u_{rms}^{2}=2\int_{\Lambda_{f}}^{\Lambda_{0}}E(k)dk\approx \frac{1}{\alpha-1}(\Lambda^{-\alpha+1}_{f}-\Lambda^{-\alpha+1}_{0})
\end{equation}

\noindent    Here, $\Lambda_{0}$ and $\Lambda_{f}$ are  the  u.v.  and i.r. cut-offs.  In principle, the cut-off $\Lambda_{0}$  can be chosen as 
 Kolmogorov's  dissipation scale.  In this case,
$\Lambda_{0}\approx (\frac{{\cal E}}{\nu^{3}})^{\frac{1}{4}}$. 
As $\nu\rightarrow 0$, $\Lambda_{0}\rightarrow \infty$ and the ratio $\Lambda_{0}/\Lambda_{f}\rightarrow \infty$ simultaneously with $Re_{T}$. Therefore,  dimensionless parameter $Re_{2,0}=\Lambda_{0}/\Lambda_{f}$
can also be regarded as a Reynolds number.  In the opposite limit  $Re_{2,0}\rightarrow 1$, the parameter $Re_{T}\rightarrow 0$.  The total  rms velocity is
$(u^{t}_{rms})^{2}=(u_{0,rms})^{2}+u_{rms}^{2}$, so that as $\Lambda\rightarrow \Lambda_{f}$, $Re\rightarrow Re_{tr}$.
This property will be important below.\\


 \noindent  
 {\it The renormalization group.} The renormalization group for fluid flows has been developed in Refs.[3]-[4] and was generalized to enable computations of various dimensionless amplitudes in the low order in the $\epsilon$-expansion in Refs.[5]-[7]. 

\noindent  Introducing velocity and length scales  $U=\sqrt{D_{0}/(\nu_{0}\Lambda_{0}^{2})}$ and  $X=1/\Lambda_{0}$, respectively, the equation (2) can be written as (for simplicity we do not change notations  for dimensionless variables):

$$
\frac{\partial {\bf u}}{\partial T}+\hat{\lambda}_{0}{\bf u\cdot \nabla u}=-\hat{\lambda}_{0}{\nabla}p +\nabla^{2}{\bf u}+ \frac{{\bf f+F}}{\sqrt{D_{0}\nu_{0}\Lambda^{2}}}
$$

\noindent where the dimensionless coupling constant (``bare'' Reynolds number) is:  $\hat{\lambda_{0}}^{2}=\frac{D_{0}}{\nu_{0}^{3}\Lambda_{0}^{\epsilon}}$
We start with the Reynolds number $Re_{\lambda}\rightarrow \infty$ and iteratively eliminating the small-scale fast modes  project original equation onto a smaller wave-number space of the  large-scale modes. \\


\noindent     Unlike renormalization group theories developed for infinite fluids in this work,  trying to accommodate information about transition and other finite-size effects, we keep the integral scale $L=1/\Lambda_{f}=O(1)$.  This is the major new element of the model. Technical  details of  all calculations presented below are best described in Ref. 6. 
Formally introducing   modes $u^{<}({\bf k},t)$ and $u^{>}({\bf k},t)$ with $k$ from the intervals $k\leq \Lambda_{0} e^{-r}$  and 
$\Lambda_{0}^{-r}\leq k\leq \Lambda_{0}$, respectively,  and averaging  over small-scale fluctuations, leads to equation for the large-scale modes [6]:

\begin{equation}
\frac{\partial {\bf u^{<}}}{\partial t}+{\bf u\cdot \nabla u^{<}}=-\nabla p +(\nu_{0}+\Delta\nu)\nabla^{2}{\bf u^{<}}+ {\bf F}+{\bf f}+{\bf \Delta f} +HOT
\end{equation}

\noindent where for simplicity correction to ``viscosity'' is written in the wave-number space:
\begin{equation}
\Delta\nu=A_{d}\frac{D_{0}}{\nu_{0}^{2}}[\frac{e^{\epsilon r}-1}{\epsilon \Lambda^{\epsilon}_{0}}+O(\frac{k^{2}}{\Lambda_{0}^{\epsilon+2}}\frac{e^{(\epsilon+2)r}-1}{\epsilon +2}) +O(\hat{\lambda}_{0}^{4})]
\end{equation}

\noindent   $\epsilon=4+y-d$ and 
$A_{d}=\hat{A_{d}}\frac{S_{d}}{(2\pi)^{d}}; \  \hat{A_{d}}=  \frac{1}{2}\frac{d^{2}-d}{d(d+2)}$.  
Due to  Galileo invariance,  high-order  $(n>1)$  terms  (HOT) generated by scale-elimination  are of the order:

\begin{eqnarray}
HOT=[\sum_{n=2}^{\infty}\hat{\lambda_{1}}^{2n}\tau_{0}^{n-1}(\partial_{t}{\bf u^{<}}+{\bf u^{<}\cdot \nabla)^{n}]u^{<}} +
O(\hat{\lambda}_{0}^{4}\nabla S^{2}_{ij}\frac{1}{\Lambda^{2}_{0}}\frac{e^{(\epsilon+2)r}-1}{\epsilon+2}) +\cdot\cdot\cdot
\end{eqnarray}

\noindent with   $\tau_{0}\approx 1/(\nu_{0}\Lambda^{2}_{0})$ and $\hat{\lambda}_{1}=\hat{\lambda}_{0}(e^{\epsilon r}-1)$.  In addition, the expression  (6) includes various products of time- and space-derivatives responsible, for example, for the rapid distortion effects (RDE).  The high-order nonlinearities generated by the procedure are small if the eliminated shell is very thin but, as will be shown below, they exponentially grow with increase of    $r$.

\noindent  {\bf As long as $k\ll \Lambda_{0}$, the $O(k^{2})$ contributions to correction to bare viscosity can be neglected together with the high-order nonlinearities. However, below,  by iterating the procedure, 
we eliminate all modes with $k\geq \Lambda_{f}$ and therefore the $O(k^{2})$ terms in (6) must be treated with care, especially in the interval $k\approx \Lambda_{f}$.}
In the mean time,  simply neglecting them, gives:

\begin{equation}
\nu_{1}(r)=\nu_{0}(1+A_{d}\hat{\lambda}_{0}^{2}\frac{e^{\epsilon r}-1}{\epsilon})
\end{equation}

\noindent  The induced noise ${\bf \Delta f}$ will be analyzed below. 

\noindent Next, starting with the equations  (6) defined on the   interval $k<\Lambda_{o}e^{-r}$, we can eliminate the  modes from the next shell of wave-numbers  $\Lambda_{0}e^{-2r}\leq k\leq \Lambda_{0}e^{-r}$ and derive  equations  of motion with another  set of corrected  transport coefficients.  The procedure can be iterated resulting in  the cut-off-dependent  viscosity, induced force etc. Setting $r\rightarrow 0$
leads  to differential recursion  equations:  denoting $\Lambda(r)=\Lambda_{0} e^{-r}$, one obtains:

\begin{eqnarray}
\frac{d\nu(r)}{d r}=A_{d}\nu(r)\hat{\lambda}^{2}(r); \hspace{2cm} \hat{\lambda}^{2}(r)=\frac{D_{0}}{\nu^{3}(r)\Lambda^{\epsilon}(r)}\nonumber 
\end{eqnarray}

 \noindent and:
 
 \begin{eqnarray}
 \frac{d\hat{\lambda}^{2}(r)}{dr}=\hat{\lambda}^{2}(r)[\epsilon-3A_{d}\hat{\lambda}^{2}(r)]\nonumber
 \end{eqnarray}
 
 \noindent Thus:
 
 \begin{eqnarray}
 \nu(r)=\nu_{o}[1+\frac{3}{\epsilon}A_{d}\hat{\lambda}_{0}^{2}(e^{\epsilon r}-1)]^{\frac{1}{3}}=
  \nu_{0}[1+\frac{3A_{d}}{\epsilon \nu_{0}^{3}}\frac{D_{0}S_{d}}{(2\pi)^{d}}(\frac{1}{\Lambda^{\epsilon}(r)}-\frac{1}{\Lambda_{0}^{\epsilon}})]^{\frac{1}{3}}\nonumber
 \end{eqnarray}
 
 \begin{eqnarray}
  \hat{\lambda}(r)\equiv[\frac{D_{0} S_{d}}{(2\pi)^{d}\nu^{3}(r)\Lambda^{\epsilon}(r)}]^{\frac{1}{2}}=
  \hat{\lambda}_{0}e^{\epsilon r/ 2}[1+\frac{3}{\epsilon}A_{d}\hat{\lambda}_{0}^{2}(e^{\epsilon r}-1)]^{-\frac{1}{2}} =
 \hat{\lambda}_{0}e^{\frac{\epsilon r}{2}}[1+\frac{3A_{d}}{\epsilon \nu_{0}^{3}}\frac{D_{0} S_{d}}{(2\pi)^{d}}(\frac{1}{\Lambda^{\epsilon}(r)}-\frac{1}{\Lambda_{0}^{\epsilon}})]^{-\frac{1}{2}}
 \end{eqnarray}
 
 \noindent and the solution for the ``induced''c  coupling  coupling constant $\hat{\lambda}_{1}$ we have:
 
 \begin{equation}
 \hat{\lambda}_{1}(r)=\frac{\sqrt{\epsilon}e^{\frac{\epsilon r}{2}}}{\sqrt{\frac{\epsilon}{\hat{\lambda}_{1}^{2}(0)}+3A_{d}(e^{\frac{\epsilon r}{2}}-1)}}
 \end{equation}

\noindent  The fixed -point- dimensionless coupling constant:

\begin{equation}
\hat{\lambda}^{*}=(\frac{\epsilon}{3A_{d}})^{\frac{1}{2}} \approx 1.29\sqrt{\epsilon}\approx 2.58.
\end{equation} 
 
\noindent   for $\epsilon=4$.  We  also see that, even when $\hat{\lambda}_{1}(0)=0$, at the fixed point $\hat{\lambda}^{*}\approx \hat{\lambda}_{1}^{*}$.

{\it Parameters.} All calculations presented below,  made in the lowest  order  of expansion in powers of  $\hat{\lambda}^{*}$  ( $\epsilon$-expansion), are described in great  detail in Ref.[6].  Eliminating all modes from the interval $k\geq \Lambda_{0}e^{-r}$ and setting $\epsilon=4$  gives:

\begin{eqnarray}
\nu(k)=(\frac{3}{8}A_{d}2D_{0})^{\frac{1}{3}}k^{-\frac{4}{3}}\approx 0.42(\frac{2D_{0}S_{d}}{(2\pi)^{d}})^{\frac{1}{3}}k^{-\frac{4}{3}}\nonumber
\end{eqnarray}

\noindent and from the linearized  equation   at the fixed point:

\begin{eqnarray}
{\bf u}^{<}\approx G(k,\omega){\bf f}=\frac{{\bf f +F(\Lambda_{f})}}{-i\omega+\nu(k)k^{2}}\noindent 
\end{eqnarray}

\noindent we derive Kolmogorov's spectrum valid in the range $k>\Lambda_{f}$:

\begin{eqnarray}
E(k)=\frac{1}{2}\frac{S_{d}k^{2}}{(2\pi)^{d+1}}\int_{-\infty}^{\infty}TrV_{ij}({\bf k}\omega)d\omega=
\frac{1}{2(\frac{3}{8}\hat{A}_{d})^{\frac{1}{3}}}(2D\frac{S_{d}}{(2\pi)^{d}})^{\frac{2}{3}}k^{-\frac{5}{3}} =
1.186(2D_{0}\frac{S_{d}}{(2\pi)^{d}})^{\frac{2}{3}}k^{-\frac{5}{3}}
\end{eqnarray}

\noindent where $(2\pi)^{d+1}V_{ij}({\bf k},\omega)=\frac{u^{<}_{i}({\bf k},\omega)u^{<}_{j}({\bf k'},\omega')}{\delta{\bf k+k'})\delta(\omega+\omega')}$.  In the so called EDQNM  approximation, which is exact at the Gaussian fixed point,    the force amplitude $D_{0}$ can be related to the  mean dissipation rate [6], [13], [14]]:

\begin{equation}
2D_{0}S_{d}/(2\pi)^{d}\approx 1.59{\cal E}; \hspace{2cm} E(k)=C_{K}{\cal E}^{\frac{2}{3}}k^{-\frac{5}{3}}; \hspace{2cm} C_{K}=1.61
\end{equation}

\noindent  Let us identify  the  infra-red cut - off $\Lambda_{f}=\Lambda(r)\approx 1/L$  with the wave-number corresponding to  the top of the inertial range. In the large Re-limit $\Lambda_{0}/\Lambda_{f}\gg1$, the total energy of the inertial range turbulent fluctuations is evaluated readily:

\begin{eqnarray}
{\cal K}=\int_{\Lambda_{f}}^\infty E(k)dk=\frac{3}{2}C_{K}(\frac{{\cal E}}{\Lambda_{f}})^{\frac{2}{3}}=
\frac{3}{2}1.61(\frac{3}{8}\hat{A}_{d}1.59)^{\frac{1}{3}}\frac{{\cal  E}}{\nu(\Lambda_{f})\Lambda_{f}^{2}}\approx 
1.19\frac{{\cal  E}}{\nu(\Lambda_{f})\Lambda_{f}^{2}}
\end{eqnarray}

\noindent and,  setting $k=\Lambda_{f}$   gives the  expression for effective viscosity in  equation for the large-scale dynamics in the interval of scales $k<\Lambda_{f}$:

\begin{equation}
\nu_{T}\equiv  \nu(\Lambda_{f})\approx 0.084\frac{{\cal K}^{2}}{{\cal E}};  \hspace{1cm} 10.0\times \nu(\Lambda_{f})^{2}\Lambda_{f}^{2}={\cal K}
\end{equation}\\
 
{ \it Fixed- point Reynolds number and irrelevant variables.}    The expression (16) gives effective viscosity accounting for all turbulent fluctuations from the interval $1/\Lambda_{0}\leq r<L=1/\Lambda_{f}$ acting on  the almost-coherent-large scale flow on the scales $r\approx L=1/\Lambda_{f}$.  Using (15) -(16):
$R_{\lambda,fp}={2\cal K}\sqrt{5/(3{\cal E}\nu(\Lambda_{f}))}=\sqrt{20/(3\times 0.084)} =9.0$. 
The same parameter   can be expressed in terms of the fixed-point coupling constant:

 

\begin{eqnarray} 
\hat{\lambda}^{*}=\sqrt{\frac{D_{0}S_{d}/(2\pi)^{d}}{\nu_{T}^{3}\Lambda_{f}^{4}}}=
\sqrt{\frac{0.8{\cal E}}{\nu_{T}^{3}\Lambda_{f}^{4}}}=\frac{\sqrt{0.8\times 400{\cal E}\nu_{T}}}{u_{rms}^{2}}
=\frac{\sqrt{0.8\times 400\times \frac{5}{3}}}{R_{\lambda}^{fp}}=\sqrt{\frac{4}{3\hat{A}_{d}}} =2.58\nonumber
\end{eqnarray}

\noindent and  $R_{\lambda,fp}\approx 9.0$ very close to  Reynolds number of transition $R_{\lambda}\approx 9-10$,   
obtained from direct numerical simulations of Ref.[11].   This outcome  correlates  with an observation that in the flows past various bluff bodies, 
the Reynolds number based on the measured ``turbulent viscosity'', integral length-scale and large-scale rms velocity  is $R_{\lambda,T}=O(10)$,  independent on the ``bare''  (classic) Reynolds number calculated with molecular viscosity [15].  

\noindent   

 In derivation of parameters (9)-(11)  all $O(k^{4})$ contributions to the propagator and  non-linearities ($HOT$ )  remained undetermined and were neglected.    As  follows from relation (10), each term of the expansion is not small:  indeed, the initially negligible small coupling constant $\hat{\lambda}_{1}$ exponentially  grows to the $O(1)$ value at the fixed point. Still, the  procedure   led to various dimensionless amplitudes in a surprisingly good agreement with experimental data. In addition,  it enabled   derivation  of  turbulence models widely 
 used in engineering [3]-[9].    The reasons for this ``surprise'' can be understood as follows. 
{\it 1. \ The above calculations gave  for viscosity at the integral scale   $ \nu^{fp}=\nu(\Lambda_{f})=\nu_{tr}$}.  2.  \  As was shown numerically in Ref.[11],  transition to turbulence  is ``smooth'' meaning that neither velocity nor  it's  spatial derivative are discontinuous at transition point. Therefore,  at the integral scale:

\begin{eqnarray}
\frac{D{\bf u}_{0}}{Dt}=-\nabla p+\nu_{tr}\nabla^{2}{\bf u}_{0}+{\bf F}(\Lambda_{f});\nonumber \\
 \frac{D{\bf u}^{fp}}{Dt}=-\nabla p+\nu^{fp}\nabla^{2}{\bf u}^{fp}+{\bf F}(\Lambda_{f})+\psi +HOT
\end{eqnarray}

\noindent where $\frac{D}{dt}\equiv \partial_{t}+{\bf u\cdot} \nabla$ and $\psi$ is induced  gaussian ``eddy noise'' (backscattering).  The field  $u_{0}$ describes a coherent,  slowly- varying or stationary  transitional flow and  the time /ensemble - averaged noise $\overline{\psi}=0$ (see below).  Since  $\nu_{tr}=\nu^{fp}$ and smoothness of transition means ${\bf u}^{fp}={\bf u}_{0}$,  it follows from (17): 
{\bf  at the integral scale  (fixed point)
 all high-order non-linearities,  generated by the coarse-graining   disappear, i.e. 
$HOT=0$}.  \\


\noindent  A  connection of the present work to Landau's theory of transition to turbulence [10] is possible. 
Let us consider  the linearized equation of motion in the vicinity of the fixed point where   ${\bf u}={\bf u_{0}}+{\bf u}_{1}$:

\begin{equation}
\frac{\partial {\bf u}_{1}}{\partial t}+{\bf u}_{0}\cdot \nabla {\bf u}_{1}+{\bf u}\cdot \nabla {\bf u}_{0}=-\nabla p_{1}+\nu\nabla^{2}{\bf u}_{1}+HOT
\end{equation} 

\noindent  If the  first unstable mode ${\bf u}_{1}\propto Ae^{i\omega}$, then according to Landau's theory:  $u_{1}\propto  A_{max}\propto \sqrt{Re-Re_{tr}}$


\noindent  and all nonlinearites are calculated from the balance:

$$HOT\approx{\bf u}_{0}\cdot \nabla {\bf u}_{1} \approx u_{0}\Lambda_{f}\sqrt{Re-Re_{tr}}$$

{\it Large-scale dynamics.} Now we would like to discuss the large-scale flow in the interval $k\approx \Lambda_{f}$, where  the bare force ${\bf f}({\bf k})=0$ and therefore the equation of motion is:

\begin{eqnarray}
\frac{\partial {\bf u^{<}}}{\partial t}+{\bf u^{<}\cdot \nabla u^{<}}=-{\nabla}p +
 \nu(\Lambda_{f})\nabla^{2}{\bf u^{<}}+ {F+\bf \psi}
\end{eqnarray}

\noindent with induced noise evaluated in Ref.[6]:

\begin{equation}
\overline{\psi_{i}({\bf k})\omega)\psi_{j}({\bf k'},\omega')}= (2\pi)^{d+1}2D_{L}k^{2}P_{ij}({\bf k})\delta({\bf k+k'})\delta(\omega+\omega')
\end{equation}

\noindent where 

\begin{equation}
D_{L}=D_{0}\frac{d^{2}-2}{20d(d+2)}\frac{\hat{\lambda}^{*}\hat{\lambda}^{*}}{\Lambda_{f}^{5}}=D_{0}\frac{0.155}{\Lambda_{f}^{5}}
\end{equation}

\noindent The induced force $\psi$ is the result of  small-scale turbulent fluctuations on the large-scale dynamics which is often called ``backscattering''. 
In the most important limit $k\rightarrow \Lambda_{f}$ 
$
D_{0}/D_{L}\approx 1/0.155 \approx 7.0$.  Therefore, the induced force, while being  numerically   not too  large,  is responsible for both 
blurring of the  large-scale transitional patterns and for the observed   
gaussian statistics  of the large-scale velocity fluctuations  in the high-Reynolds number flows . \\

\noindent {\it Summary and conclusions.}     1. \ The coarse-graining procedure based on  Wyld's diagrammatic expansion leads to the Navier-Stokes-like  equations with an   infinite number of additional higher -order nonlinearities. In the lowest -order of expansion,  the calculated fixed-point Reynolds number $Re_{fp}\approx Re_{tr}$ where $Re_{tr}$ is  a numerically computed Reynolds number of transition to turbulence.  Since  the numerically discovered transition is ``smooth'',  {\bf we assumed} that  at the transition point both the velocity fields ${\bf u}_{0}={\bf u}_{fp}$ and their spatial derivative are equal  ($\nabla_{i}u_{0,j}=\nabla_{i}u_{fp,j}$)  and,  as a result,   at the fixed point all additional to the NS equations high-order nonlinearities are irrelevant.  2. \   Comparison with Landau's  theory of transition to turbulence shows that the nonlinear terms are $O(\sqrt{Re-Re_{tr}})\rightarrow 0$. 
3. The infra-red divergencies appearing in the each term of the expansion do not disappear but are summed up into equations of motion for the large-scale features of the flow.\\

\noindent Previous  theories of turbulence used all sorts of field-theoretical approaches   to entirely remove infra-red divergencies.  In this work we show that any large-Reynolds-number  flow includes  the symmetry-breaking large-scale field responsible for  the small-scale turbulence production. The i.r. divergencies of turbulence theory do not disappear but are contained  in the equations  of motion for the 
large-scale quasi-laminar (coherent) component reflecting geometry, physical mechanisms of production etc. The symmetries of turbulent flow like a universal behavior of the moments of derivatives and structure functions are  recovered at somewhat  smaller  scales. This result probably has a wide  range of applicability.  For example, Polyakov, in his conformal field theory of   {\it small-scale dynamics}  of two-dimensional turbulence  attributed the infra-red problems appearing in the vicinity of  the forcing scale to  ( ``non-conformist''  ) symmetry-breaking ''condensates'',  i.e. large-scale flows generating ``direct''  enstrophy cascade [16]. \\

\noindent The author is grateful to A. M. Polyakov and I. Staroselsky for their interest in this work and  numerous suggestions.\\

\noindent {\it References.}\\
\noindent 1.  H.W. Wyld,   
Annals of Physics {\bf 14}, 143-165 (1961);\\
\noindent 2. R.H.Ktaichnan, Phys. Fluids {\bf 7}, 1723 (1964);\\
\noindent 3. \ D. Forster, D. Nelson \&  M.J. Stephen, Phys.Rev.A {\bf 16}, 732 (1977);\\
\noindent 4. \ C. DeDominisis \& P.C. Martin, Phys.Rev.A{\bf 19}, 419 (1979);\\
\noindent 5. \  
 V. Yakhot \&  S.A. Orszag, Phys.Rev.Lett.{\bf 57}, 1722 (1986);\\
\noindent 6.  \ V. Yakhot \& L. Smith, 
J. Sci.Comp. {\bf 7}, 35 (1992);\\
\noindent 7.  \ V. Yakhot, S.A. Orszag, T. Gatski, S. Thangam \& C.Speciale, 
 Phys. Fluids A{\bf 4}, 1510  (1992);\\
 \noindent 8. H. Chen,   S. Kandasamy,  S.A. Orszag,   R. Schock,
S. Succi, \&  Yakhot V. 2003,  
Science {\bf 301}, 633--63;\\
 \noindent  9. \  C. Bartlett, H. Chen, I. Staroselskii \& V.Yakhot,   Int.Journ. of Heat and Fluid Flow {\bf 42}, 1 (2013);  pp 235-374;\\
 \noindent 10. \ L.D.Landau \& E.M. Lifshits, ``Fluid Mechanics'', Pergamon, New York, 1982;\\
\noindent 11. \  J.Schumacher, K.R. Sreenivasan \& V. Yakhot; 
New Journ. of Physics  {\bf 9}, 89 (2007); \\
\noindent 12. \    \  S.C.C. Bailey,  M.Hultmark, A.J. Smits, J. Schumacher, V. Yakhot, 
Phys. Rev. Lett., {\bf 103} 014502 (2009); \\
\noindent 13. \  S.A. Orszag,  in ``Fluid Dynamics''  edited by R. Balian and J.-L Puebe (Gordon and Breach, New York 1977) pp. 235-374;\\
\noindent 14. \ W.P. Dannevik,  V. Yakhot \& S.A.Orszag, Phys.Fluids {\bf 30}, 2021 (1987);\\
\noindent 15. \  K.R. Sreenivasan, Private communication (2013);\\
\noindent 16. \ A.M. Polyakov, Nucl.Phys. B{\bf 2}, 393 (1993); \\

\end{document}